# Tunable resonant Raman scattering with temperature in vertically aligned 2H-SnS$_2$

Atul G. Chakkar[1, *], Deepu Kumar[1], Ashok Kumar[2], Mahesh Kumar[2], and Pradeep Kumar[1, #]

[1]*School of Physical Sciences, Indian Institute of Technology Mandi, Mandi-175005, India*
[2]*Department of Electrical Engineering, Indian Institute of Technology Jodhpur, 342037, India*

[*]E-mail: atulchakkar16@gmail.com
[#]E-mail: pkumar@iitmandi.ac.in

**Abstract**

Two-dimensional semiconducting materials have a wide range of applications in various fields due to their excellent properties and rich physics. Here, we report a detailed investigation of the temperature dependent Raman and Photoluminescence measurements on the vertically aligned 2H-SnS$_2$ grown by CVD method. Our results established the tunability of the resonant Raman scattering with varying temperature, i.e. a crossover between resonance and non-resonance conditions for the current system. We also discussed the temperature as well as laser power dependence of the low frequency asymmetric Raman mode which is interlayer shear mode. Temperature dependence of the intensity of the phonon modes also manifests the tunability of the resonant Raman scattering with temperature. Our temperature dependent Photoluminescence measurement shows the strong temperature dependence of the excitonic peaks which is confirmed with laser power dependance of the Photoluminescence measurement at room temperature. Our investigation may help to design and fabricate devices based on vertically aligned 2H-SnS$_2$ and other similar materials in future.

**Keywords:** 2D semiconductor, vertically aligned, SnS$_2$, Raman, photoluminescence, resonance effect,

1. **Introduction**

Recently, two-dimensional (2D) layered semiconducting transition metal dichalcogenide materials gaining more research interest due to their unique properties mainly intrinsic large band gap, and excitonic features and possible potential applications in gas sensing, optoelectronics, energy storage, and electronic devices [1–3]. Among the 2D materials family, the vertically aligned (VA) tin disulfide $SnS_2$ in 2H phase has been emerged as a potential candidate for chemical gas sensing, optoelectronics, photodetectors, and other device applications due to its large surface area-to-volume ratio, high transistor switching ratio, fast photo-detection, high electrical conductivity, wide bandgap, and excitonic feature [4–8]. Additionally, it is also cost-effective, low in toxicity and high in stability. $SnS_2$ is a polar semiconductor with band gap $E_g$ ~ 2.3-2.5 eV [9], and hence it may provide a crucial path to study excitonic features.

In the resonant as well as non-resonant Raman scattering process, these excitonic energy states in semiconducting materials play a crucial role. A study of the thermal properties of such potential semiconducting candidates explores more properties which may be helpful in various applications such as thermoelectric power generation and in device applications like transistors and sensors [10,11]. Thermodynamic properties such as thermal expansion, thermal conductivity, and specific heat are strongly affected by the anharmonicity effect and other underlying phenomena. Due to anharmonicity, the phonon has a finite lifetime and reduces self-energy. Anharmonicity strongly influences the thermal transport, electronic, and optoelectronic properties of materials. In layered 2D transition metal dichalcogenides (TMDs) and other materials, the electronic, optoelectronic, and thermal properties are associated with anharmonic effects and to understand the anharmonicity and phonon dynamics in such materials, a temperature-dependent Raman spectroscopic investigation is important [12–22]. Raman spectroscopy is one of the most promising non-destructive tools to understand the thermal, structural, electrical, magnetic, and optical properties of 2D materials [13,18,23–34].

In this work, we have performed detailed temperature dependent Raman and Photoluminescence (PL) measurements in the temperature range of ~ 10-310 and 5-320 K, respectively and analyzed to understand the dynamics of the phonon modes and the exciton, respectively as a function of temperature. We also performed power dependent Raman measurement in the low frequency regime and power dependent PL measurement with 325 nm laser excitation to investigate the excitonic nature of the observed PL peaks.

2. **Experimental details**

We have performed the temperature dependent Raman and PL measurement on vertically aligned (VA) 2H-$SnS_2$ flakes grown by chemical vapor deposition (CVD) method [35]. Both Raman and PL measurements were performed using a Horiba Lab-Ram HR Evolution Raman spectrometer in backscattering configuration, while temperature changed using a closed cycle refrigerator (Montana cryostat) in the temperature range of 10-310 K, with a temperature accuracy of ±0.1 K. For temperature-dependent Raman measurements, we used a laser excitation wavelength of 532 nm (2.33 eV), and kept laser power on sample very low, ≤ 0.2 mW, to avoid local heating effect. We recorded a Raman spectrum using a 50x long working objective lens to focus incident light with grating 1800 groves/mm. We also performed power dependent Raman measurement in the low frequency regime. For PL measurements, we used a laser excitation wavelength of 325 nm (3.81 eV), and PL recorded using a 15x NUV objective lens to focus incident light on the sample with grating 600 groves/mm. We also performed power dependent PL measurement at room temperature to understand the excitonic nature of the observed PL peaks.

## 3. Results and Discussions

### 3.1. Lattice Vibrations in VA SnS$_2$

Bulk 2H-SnS$_2$ belongs to the point group $D_{3d}^3$ (space group $P\bar{3}m1, \#164$)), and has a hexagonal structure [9,35]. Bulk 2H-SnS$_2$ consists of 3 atoms per unit cell which gives rise to nine phonon branches at the $\Gamma$-point of the Brillouin zone, expressed by following irreducible representation as $\Gamma_{irred.} = A_{1g} + 2A_{2u} + E_g + 2E_{2u}$ [9,36–39]. There are 6 optical phonon branches which are divided into three Raman active ($A_{1g} + E_g$) and three infrared actives ($A_{2u} + E_u$). The remaining three branches ($A_{2u} + E_u$) correspond to the acoustic phonons. The phonon modes with symmetry $E$ are doubly degenerate and appear due to the in-plane vibration of atoms, while the phonon modes with symmetry $A$ correspond to the non-degenerate out-of-plane vibrations of atoms. Figure 1 (a) shows the Raman spectrum of SnS$_2$ in a spectral range of 10-800 cm$^{-1}$ recorded at 300 K. Insets in the yellow-shaded area are the amplified spectra for the spectral range of 10-50 cm$^{-1}$, 50-250 cm$^{-1,}$ and 450-800 cm$^{-1}$. We have fitted the spectra using a sum of Lorentzian functions in order to extract phonon self-energy parameters i.e., mode frequency (ω), full width at half maximum (FWHM), and intensity of the individual modes. In the low-frequency regime, the peak P1 observed at ~27 cm$^{-1}$ could be attributed to the interlayer shear/breathing mode which appears due to the interaction between the adjacent layers. A very strong characteristic peak P6 located at ~ 314 cm$^{-1}$, is attributed to the first-order Raman active $A_{1g}$ mode and appears due to the out-of-plane vibration of S atoms in opposite direction with fixed Sn atoms. A very weak peak P4 located near ~205 cm$^{-1}$ is attributed to the first order Raman active $E_g$ mode and originates due to the in-plane vibration of both S and Sn atoms in opposite direction. In addition to these characteristic Raman modes, we also observed a few additional modes from which some of the modes are intense at our highest recorded temperature while some of them are intense at low temperature. For example, the phonons P2 ~ 94 cm$^{-1}$ and P3 ~ 140 cm$^{-1}$ are observed to be strong at our highest recorded temperature (~310 K) while these modes vanished completely at very low temperature. However, the peak P5 ~ 215 cm$^{-1}$ is strong at low temperature, and is absent at high temperature, see Fig. 1(b). The peak P7 ~580 cm$^{-1}$ and P8 ~ 624 cm$^{-1}$ could be attributed to the second order phonons and is assigned as $E_u(TO)+A_{2u}(LO)$ and $2E_u(LO)$, respectively [37]. These two phonons are observed to be strong at high temperature compared to the low temperature. Furthermore, we also noticed a few extremely weak phonons marked by asterisk (*) and number sign (#). The self-energy parameters of these modes could not be extracted using a Lorentzian fitting function due to their very weak nature. It is worth noting that the emergence/disappearance of phonon modes as a function of temperature may be an indication of the presence of resonance effect and modulation of this effect with temperature which is a typical phenomenon in these kind of semiconducting materials will be discussed later in more detail.

### 3.2. Temperature dependence of the phonon modes
### 3.2.1. Phonon anharmonicity and thermal expansion coefficient

Figures 2 (a-d), shows the temperature dependence of the frequency ($\omega$), and linewidth (FWHM) of the P3-P8 phonon modes in the temperature range of ~ 10-310 K. We observed that a phonon softening and broadening for all of the phonons with rise in temperature with some anomalous kink/jump near ~ 220-270 K in frequency/FWHM for some of the phonons. The change in frequency of the P4, P6, P7, and P8 is observed to be ~ 4.1, 3.3, 5.3 and 7.7 cm$^{-1}$, respectively. On the other hand, the change in FWHM of P4, P6, and P8 is found to be ~ 6.6, 5.4, and 5.6 cm$^{-1}$, respectively. Surprisingly, the change in FWHM of the P7 mode is found to be significantly large (~ 57 cm$^{-1}$). It should be noted that in our recorded temperature range, the phonon modes generally soften or broaden by a value of 1-2 cm$^{-1}$. The observed large change in frequency and FWHM of the phonon modes may indicate the involvement of electron-phonon coupling which plays a crucial role in phonon self-energy parameters under resonance condition. However, extremely large change in FWHM of the P7 mode is still not clear and may be the subject of future ongoing investigations. Nevertheless, the P7 and P8 modes are second order phonon modes [37]. Generally, in 2D TMD materials, temperature-dependent frequency of phonons may be understood using contribution of: (i) anharmonic effect and (ii) thermal expansion of the lattice. The change in frequency of the phonon modes as a function of temperature considering above two effects may be gives as [40,41]:

$$\Delta\omega(T) = \Delta\omega_{anh.}(T) + \Delta\omega_{latt.}(T) \tag{1}$$

The first term in equation (1), corresponds to the change in frequency of the phonon modes due to anharmonic effect which arises from the phonon-phonon interaction via three and four phonon processes and is given as [42]:

$$\Delta\omega_{anh}(T) = \omega(T) - \omega_0 = A\left(1 + \frac{2}{(e^x - 1)}\right) + B\left(1 + \frac{3}{(e^y - 1)} + \frac{3}{(e^y - 1)^2}\right) \tag{2}$$

Where, $x = \hbar\omega_0 / 2k_B T$, and $y = \hbar\omega_0 / 3k_B T$ associated with the three and four phonon decay processes, respectively. Constants A and B are the self-energy fitting parameters associated with the three and four phonon anharmonic processes respectively. Generally, the change in linewidth of the phonon modes as the function of the temperature can be understood by considering three and four phonon anharmonic effect and is given as [42]:

$$\Delta\Gamma(T) = \Gamma(T) - \Gamma(0) = C\left(1 + \frac{2}{(e^x - 1)}\right) + D\left(1 + \frac{3}{(e^y - 1)} + \frac{3}{(e^y - 1)^2}\right) \quad (3)$$

Where, the constants C and D are the self-energy fitting parameters, which shows the strength of the phonon-phonon interaction associated with the three and four phonon anharmonic processes, respectively. In figs. 2 (a)-(d), the sold red lines are the fitted curves using equations (2) and (3). The semi-transparent blue lines are drawn for the guide to the eye. The best extracted values of the self-energy fitting parameters are listed in Table-I.

The second term in the equation (1) corresponds to the change in frequency due to the thermal expansion of the lattice and is given as:

$$\Delta\omega_{latt}(T) = \omega_0 \left\{ \exp\left(-3\gamma \int_{T_0}^{T} \alpha(T) dT\right) - 1 \right\} \quad (4)$$

Where, $\gamma$ is the Gruneisen parameter and $\alpha(T)$ is the linear thermal expansion coefficient (TEC) of the phonon modes. For convenience, the product of the phonon mode Gruneisen parameter and the TEC may be expressed by a polynomial of temperature and is given as:

$$\gamma\alpha(T) = a_0 + a_1 T + a_2 T^2 + .... \quad (5)$$

Where, $a_0$, $a_1$, and $a_2$ are the constant parameters and are obtained by fitting frequency of the corresponding phonon modes using combination of the equations (2) and (4) in the temperature range of 10-310 K. For the case of SnS$_2$, the value of the Gruneisen parameter ($\gamma$) of the phonon modes is unknown, hence for each phonon mode we have taken three different values i.e., $\gamma$=1,2,3. Figures. 3 shows the temperature dependance of the extracted thermal expansion coefficient ($\alpha(T)$) in the temperature range of ~ 10-310 K. For the phonon modes P4 and P6, the TEC ($\alpha(T)$

) increases with increasing temperature up to ~ 240 K and decreases with further increase in temperature. This may result due to the crossover between resonance and non-resonance effects. In 2D materials, the study of thermal properties such as anharmonic effect, thermal expansion, etc., are most important because of the wide range of applications in the device field. A detailed investigation of such thermal properties may provide crucial understanding and may provide useful information for device fabrication and other applications.

### 3.2.2. Temperature dependent intensity of the phonon modes

The temperature dependent intensity of the phonon modes in 2D semiconducting materials may provide crucial information/understanding about its electronic and optical properties [43–46]. In this section, we focus on the temperature dependent intensity of the phonons. Figures 4 (a) and (b) show the temperature dependent intensity for some of the prominent phonon modes i.e. P4, P6 and P7, P8; respectively. The intensity of P4, P6, and P8 increase gradually with an increase in temperature and attains a maximum around 200-220 K. Interestingly, the intensity is observed to decrease with a further rise in temperature above 200-220 K. Similarly, the intensity of the P7 mode attains a maxima around ~240-260 K. Such temperature-dependent intensity of the phonon modes suggests the involvement of the intriguing mechanism.

Generally, the temperature-dependent Raman scattering intensity of the phonon modes under the non-resonant condition, could be described using the Bose-Einstein function. For the Stokes Raman scattering case, the intensity of the of the first-order phonon (first-order Raman scattering), second-order sum of two phonons (creation of two phonons; second-order Raman scattering) and second order difference of two phonons (creation of one phonons while annihilation of second phonon; second-order Raman scattering) is given as $(n_1+1)$, $(n_1+1)(n_2+1)$, and $(n_1+1)(n_2)$; respectively, where n is the Bose-Einstein factor and given as $n = [\exp(\hbar\omega/\kappa_B T)-1]^{-1}$ [47]. It is worth noting that the Bose-Einstein factor is expected to increase with increase in temperature and so the intensity of the phonon modes is also expected to increase. Our observation clearly suggests that alonee Bose-Einstein expression is not sufficient to understand the temperature dependent intensity of the phonon as shown in Fig. 4.

It is worth noting that the quantum mechanical picture has played a crucial role to understand the temperature dependence intensity of the phonons, especially in the vicinity of resonance effect, in semiconducting materials [45,48]. Moreover, the excitonic energy states show a strong dependence on the temperature. This indicates the crossover in non-resonance and resonance conditions can be obtained not only by varying the laser excitation energy but also could be achieved with varying the temperature. The crossover between non-resonance and resonance conditions with varying temperature may result in the modulation in the intensity of the phonon modes. Considering quantum mechanical picture, the intensity of the phonon for first order Raman scattering process is given as [47,49]:

$$I = \left| \sum_{m,n,n'} \frac{\langle m|H_2|n'\rangle \langle n'|H_{el-ph}|n\rangle \langle n|H_1|m\rangle}{(E_L - \Delta E_{mn})(E_s - \Delta E_{mn'})} \right|^2 \tag{6}$$

Where, $|m\rangle$ $|n\rangle$, and $|n'\rangle$, represents the ground state and the intermediate states, respectively. $H_1$ and $H_2$ are the Hamiltonians representing the electron-photon (absorption) and electron-photon (emission) process, respectively. $H_{el-ph}$ is the electron-phonon interaction Hamiltonian. $\Delta E_{mn} = (E_n - E_m + i\gamma_1)$ and $\Delta E_{mn'} = (E_{n'} - E_m + i\gamma_2)$ represents the energy difference between the intermediate state $E_n$ and the ground state $E_m$, and the intermediate state $E_{n'}$ and the ground state $E_m$, respectively. While $\gamma_1$ and $\gamma_2$ are related to the finite lifetime of the intermediate states $|n\rangle$ and $|n'\rangle$. $E_L = \hbar\omega_L$ and $E_s = \hbar\omega_s = \hbar\omega_L - \hbar\omega_{ph}$ are corresponds to the incident and scattered photon energy, respectively and $\omega_{ph}$ is the frequency of the phonon involved in the Raman scattering process. For the case, if the intermediate states are real then $E_n - E_m$ and $E_{n'} - E_m$ could be written as excitonic energy states ($E_x$) or bandgap ($E_g$) of the materials. When the laser excitation energy is close to the excitonic energy states ($E_x$) or bandgap ($E_g$), it gives rise to resonance effect which significantly enhances the intensity of the phonon. Change in laser excitation energy results in to crossover between resonance and resonance conditions. Furthermore, we note that the excitonic energy states ($E_x$) or bandgap ($E_g$) is also dependent on the temperature suggesting that temperature is an additional parameter which could also led to the crossover between resonance and resonance conditions. In the vicinity of the excitonic resonance effect, the numerator term in equation (6) may become weak as compared to denominator and the Raman scattering intensity of the phonon mode may be written as:

$$I \propto \left| \frac{1}{\{E_x(T) - E_L + i\gamma_X(T)\}\{E_x(T) - E_L + i\gamma_x(T)\}} \right|^2 \tag{7}$$

Where, $E_X(T)$ and $\gamma_X(T)$ are the temperature-dependent transition energies and linewidth correspond to excitonic energy states, respectively. Equation (7) suggests that the resonance condition could be achieved by changing the incident excitation laser energy as well as by changing temperature. Generally, with increasing temperature, the energy of the excitonic states decreases. The change in the excitonic energy with temperature results in the crossover between non-resonance and resonance conditions, which may lead to the modulation in intensity of the phonon while keeping fixed the laser excitation energy. We note that SnS$_2$ is a semiconducting material with a band gap of $E_g \sim$ 2.3-2.5 eV. We have excited the Raman spectra with laser excitation energy of 2.33 eV, clearly reflecting that system is under the resonance effect near the room temperature. Hence, the enhancement in the intensity of the phonon modes is due to the resonance observed with the increasing temperature in the range of ~ 180-240 K. In section 3.5, we investigated and explained the temperature dependence of the PL peaks which support our results about crossover between non-resonance and resonance conditions for the current system.

### 3.3. Low frequency shear mode

In this section, we shed light on the low frequency interlayer phonon mode P1 observed close to ~ 27 cm$^{-1}$, We note that interlayer low frequency phonons in layered 2D materials, appear due to the interlayer interaction between the adjacent layers, which play an important role in investigation of wide range of aspects of materials such as strain effects, interlayer interaction strength, stacking configuration, electron-phonon coupling, and number of layers [28,32,50–52]. To the best of our knowledge, the detailed investigations of the low frequency Raman dynamics of the 2H-SnS$_2$ are still lacking so far. P.H. Tan et al., [53] uncover the Raman dynamics of such asymmetric shear mode in graphene as the function of the layers.

Figure 5 (a) shows the temperature evolution of the low frequency mode P1 mode which appears due to the in-plane (shear) or out-of-plane (breathing) interaction of the adjacent layers. We observed asymmetrical nature of the P1 mode and surprisingly the asymmetricity survives even up to our highest recorded temperature, see Figs. 1(a) and 5 (a). This low frequency Raman mode provides the direct measurement of the interlayer coupling between two adjacent layers in vdW's materials. Asymmetric nature of P1 mode may be due to quantum interference between a Raman allowed phonon and a continuum of Raman active electronic or multiphoton transitions [54]. To understand quantitatively, we have analyzed this asymmetric line shape of the P1 mode using a Breit-Wagner-Fano (BWF) function, which is given by [53,55]: $I(\omega) \propto \frac{(1+\delta/q)^2}{1+\delta^2}$, where $\delta = \dfrac{\omega - \omega_0}{\Gamma}$, and $I(\omega)$ is the Raman intensity as a function of frequency $\omega$. $\omega_0$ and $\Gamma$ are the frequency and FWHM of the bare /uncoupled phonon, respectively. $q$ is the asymmetry parameter that characterizes the coupling strength between the phonon and the continuum, quantified by the parameter $1/|q|$. In the limit $1/|q| \to \infty$; coupling is stronger and causes the peak to be more asymmetric while in the limit $1/|q| \to 0$; coupling is weak and this results into the Lorentzian line shape [29,56]. The solid red lines in Fig. 5(a) are the best fit using the BWF function. Figs. 5(b) and (c) show the temperature dependence of the frequency and linewidth of the peak P1 in the temperature range of ~ 10-310 K. For mode P1, both parameters (i.e., frequency and linewidth) show the discontinuity in the temperature range of ~ 220-250 K and this discontinuity may be because of the resonance phenomena in this temperature range. Figure 5(d) shows the temperature dependence of the coupling strength ($1/|q|$) in the temperature range of 10-310 K. We note that, with lowering temperature $1/|q|$ increases slightly and it shows the weak dependence on the temperature and shows the discontinuity in the temperature range of ~220-250 K.

The Fano asymmetry may be affected by the laser power which is mainly due to the laser induced or photoexcited charge carriers results in increase in free electron plasma with increasing laser power [57,58]. We also performed the laser excitation power as well as the acquisition time (spectrum collection time) dependent Raman measurement for the low frequency region to understand the effect of laser power and acquisition time on the Fano interaction. Figures 5(e) and (f) show the evolution of the Fano peak P1 and the coupling strength ($1/|q|$) as a function of the laser power. The coupling strength increases exponentially with increasing laser power, indicating the coupling between phonon and continuum increases with laser power. This enhancement in the coupling may have occurred due to the laser induced electron plasma. Figure 5(g) shows the evolution of the peak P1 as the function of acquisition time i.e., recording time of the spectrum from 10 sec to 120 sec with increasing 10 sec intervals. Fig. 5(h) shows the coupling strength ($1/|q|$) dependence on the acquisition time of the spectrum. With increasing acquisition time, $1/|q|$ decreases linearly though change is small. This suggests there is no significant plasma induced with increasing spectrum acquisition time. As we increase the laser excitation power, which induced more electron-hole plasma and results in an increase in the coupling strength with increasing laser power. This result suggests we can control the coupling between the phonon and underlying of the continuum and/or multiphoton transitions with laser excitation power.

**3.4 Polarization dependance of the phonon modes**

To understand the polarization dependence of the Raman spectrum of VA 2H-SnS$_2$, we have performed polarization dependent Raman measurements at room temperature by rotating incident light polarization and keeping scattering light polarization fixed. Ding *et al.,* [9] performed polarization dependent study on SnS$_2$ by rotating sample and suggesting application of this material in the field of polarizing devices. VA semiconducting materials are potential candidates for applications in opto-electronic and other devices. Hence, it is important to study the polarization dependance of the phonons for the current system.

Figure 6(a) shows the polar plot for the intensity of the phonon modes P2-P4, P6, P7, and P8 at temperature 300 K. P2 and P3 are weak peaks and P7 and P8 are the second order phonon peaks. P4 is the weak peak with E$_g$ symmetry, and it is doubly degenerate while P6 is the most prominent peak with A$_g$ symmetry. The Raman tensors for the phonon modes with symmetries E$_g$ and A$_g$ are listed in supplementary material [36]. To understand the polarization dependence of the phonon modes P4 and P6, we used semi-classical approach and according to this, intensity of the phonon modes is given as $I \propto |e_s^T . R.e_i|^2$, where, T is the transpose, $e_i$ and $e_s$ are the units vectors in the direction of the incident and scattered light electric filed. R is the Raman tensor for Raman modes. The unit vectors in the direction of the incident and scattered light can be written in matrix form as; $\hat{e}_i = [\cos(\theta+\theta_0) \quad \sin(\theta+\theta_0) \quad 0]$ and $\hat{e}_s = [\cos(\theta_0) \quad \sin(\theta_0) \quad 0]$, respectively. Where $\theta$ is the angle between unit vectors of incident and scattered light and $\theta_0$ is the arbitrary angle between scattered light polarization and x-axis as shown in schematic Fig. 6(c). The intensity of the phonon with $E_g$ and $A_g$ symmetries depends on the polarizing angle ($\theta$) and associated Raman tensors and is given as:

$$I_{E_g} = |c\cos(\theta)(\cos^2(\theta_0) - \sin^2(\theta_0)) - c\sin(2\theta_0)\sin(\theta)|^2, \text{ and} \quad (8(a))$$

$$I_{E_g} = |c\sin(\theta)(\sin^2(\theta_0) - \cos^2(\theta_0)) - c\sin(2\theta_0)\cos(\theta)|^2 \quad (8(b))$$

$$I_{A_g} = a^2 \cos^2(\theta) \quad (9)$$

Without loss of generality, we consider $\theta_0 = 0$. This results in the above equations (8(a)) and (8(b)) becomes:

$$I_{E_g} = c^2 \cos^2(\theta), \text{ and} \quad (10(a))$$

$$I_{E_g} = c^2 \sin^2(\theta) \quad (10(b))$$

In figure 6(a), The solid red line shows the fitted curve by adding equations (10(a)) and (10(b)),

$$I_{E_{1g}} = c^2 \cos^2(\theta) + c^2 \sin^2(\theta) = c^2 \quad (11)$$

The polar plot for the intensity of the phonon mode P6 with $A_g$ symmetry shows the nearly circular symmetry, while according to equation (9), it should be two-fold kind of symmetry. Ding et al. [6]

performed polarization dependent Raman measurements on SnS$_2$ crystal in basal and cross-plane configuration were observed circular and two-fold symmetry for basal and cross planes, respectively for $A_g$ (~ 314 cm$^{-1}$) mode. In our measurement, we observed circular symmetry for P6 mode. In our measurement, we observed circular symmetry for P6 ($A_{1g}$) mode and this may be due to the vertically aligned SnS$_2$ grown by CVD method. M. Hulman et al., [59] also studied the polarization dependent Raman measurement on vertically and horizontally aligned MoS$_2$ and observed different polarization dependance for the mode with $E_g$ symmetry.

Figure 6(b) shows the polarization dependance of the low frequency interlayer mode P1. It shows the independence on the polarizing angle $(\theta)$. This suggests the mode P1 is the low frequency interlayer shear mode which results due to the in-plane vibrations of the adjacent layers. By rotating the incoming light electric field, our polarization analysis demonstrates the polarization dependence of the phonon modes, primarily the low frequency mode P1 and the high frequency modes P4 and P6 in VA SnS$_2$.

### 3.5: Temperature and laser power dependence of the photoluminescence

To understand the dynamics of excitonic quasiparticles, the temperature dependent PL spectroscopy has been widely used [60–66]. To know the nature of excitonic-quasi-particles in 2D materials, laser excitation power dependent PL measurements are also reported for other 2D materials [67]. In 2D semiconducting materials, the coupling of excitonic quasiparticles (exciton, trions, bi-exciton) with phonons and charge carriers plays an important role in controlling the optical properties. Burton *et al.*, [68] studied the optical and electronic properties of the single crystal SnS$_2$. To the best of our knowledge, there is no detailed temperature dependent PL spectroscopic study reported for the few layered CVD grown VA 2H-SnS$_2$. Hence, we performed and analyzed temperature as well as power dependent PL measurements on VA 2H-SnS$_2$. Figure 7(a) shows the evolution of the fitted PL spectra at four different temperatures 5, 150, 230, and 300 K (for detailed temperature evolution of the PL spectra, see fig. S2). PL spectra fitted with sum of the Voigt function to extract the peak position, FWHM and intensity of the individual PL peaks at each recorded temperature and power dependent PL measurements. At low temperature, we fitted PL spectra with four individual peaks namely S1-S4. As we increase the temperature, peak S1 and S2 become weak and vanished after a certain temperature. Peak S3 and S4 show the strong temperature dependance in the temperature range of ~ 5-320 K. Figures 7(b) and (c) show the temperature dependence of the extracted parameters in the temperature range of ~ 5-320 K. For peak S3, peak position gradually decreases with increasing temperature, while FWHM increases gradually up to ~ 230 K and decreases with further increase in temperature. The position of the peak S3 approaches energy ~ 2.33 eV in the temperature range of ~200-260 K which is the energy of the laser excitation 532 nm (2.33 eV) we used for the temperature dependent Raman measurement. This suggests a strong enhancement in the resonant Raman scattering in this temperature range. For the peak S4, peak position decreases slightly with an increase in temperature up to ~ 90 K. With further increase in temperature, it increases up to ~ 230 K and decreases drastically with further increase in temperature. The FWHM of the peak S4 increases slightly with an increasing temperature up to ~ 170 K and increases drastically with further increase in temperature. PL intensity of both beaks S3 and S4 slightly increases with an increase in temperature till ~180 K and increases exponentially with further increase in temperature. Such increase in the intensity of the PL peak with increasing temperature also observed for few layer MoSe$_2$ [69] . To know the nature of the PL peaks, we also performed power dependent PL measurement at room temperature. Figure 8(a) shows the evolution of PL spectra as a function of the laser power, and it mainly shows the peak S3 and S4. Figure 8(b) shows the integrated PL intensity of the peaks S3 and S4 as a function of the laser power. With increasing laser power intensity of both peaks increases and shows the strong dependance. We fitted the intensities with power law fit: $I \propto P^{\alpha}$. The extracted value of the exponent $\alpha$ for peaks S3 and S4 are 1.02±0.06 and 0.82±0.05; respectively. These values of $\alpha$ suggest the excitonic nature of both peaks S3 and S4 [70]. D. Y. Lin et al., [71] performed temperature dependent piezoreflectance measurement on layered ternary $SnS_{2-x}Se_x$ and they also observed similar behavior for the excitonic peak around ~ 2.3 eV.

**Conclusion:**

In conclusion, we performed detailed inelastic light (Raman) scattering and photoluminescence measurements and analysis on vertically aligned 2D 2H-$SnS_2$ grown by CVD method. Our temperature dependent Raman measurements reveal the temperature dependent phonon-phonon interaction, thermal expansion coefficient and intensity of the observed phonon modes. We also performed polarization measurement to understand the symmetry of the phonon modes. Our temperature as well as laser power dependent PL measurements provide interesting results about dynamics of the observed excitonic peaks. Our results suggest the tunability (crossover) between the resonant and non-resonant Raman scattering with temperature for 2H-$SnS_2$ and our detailed investigation provides crucial understanding which may help for the design and fabrication of the future opto-electronic devices based on the VA semiconducting $SnS_2$ and other such materials.

**Acknowledgements**

P.K. thanks SERB (Project no. CRG/2023/002069) for the financial support and IIT Mandi for the experimental facilities.

**Data availabilty statement**

All data that supports the findings of this study are included within the article and supplementary file.

ACS Nano **14**, 14579 (2020).

[71] D. Y. Lin, H. P. Hsu et al., *Temperature dependent excitonic transition energy and enhanced electron-phonon coupling in layered ternary $SnS_{2-x}Se_x$ semiconductor with fully tunable stoichiometry,* Molecules **26**, 2184 (2021).24

Figures:

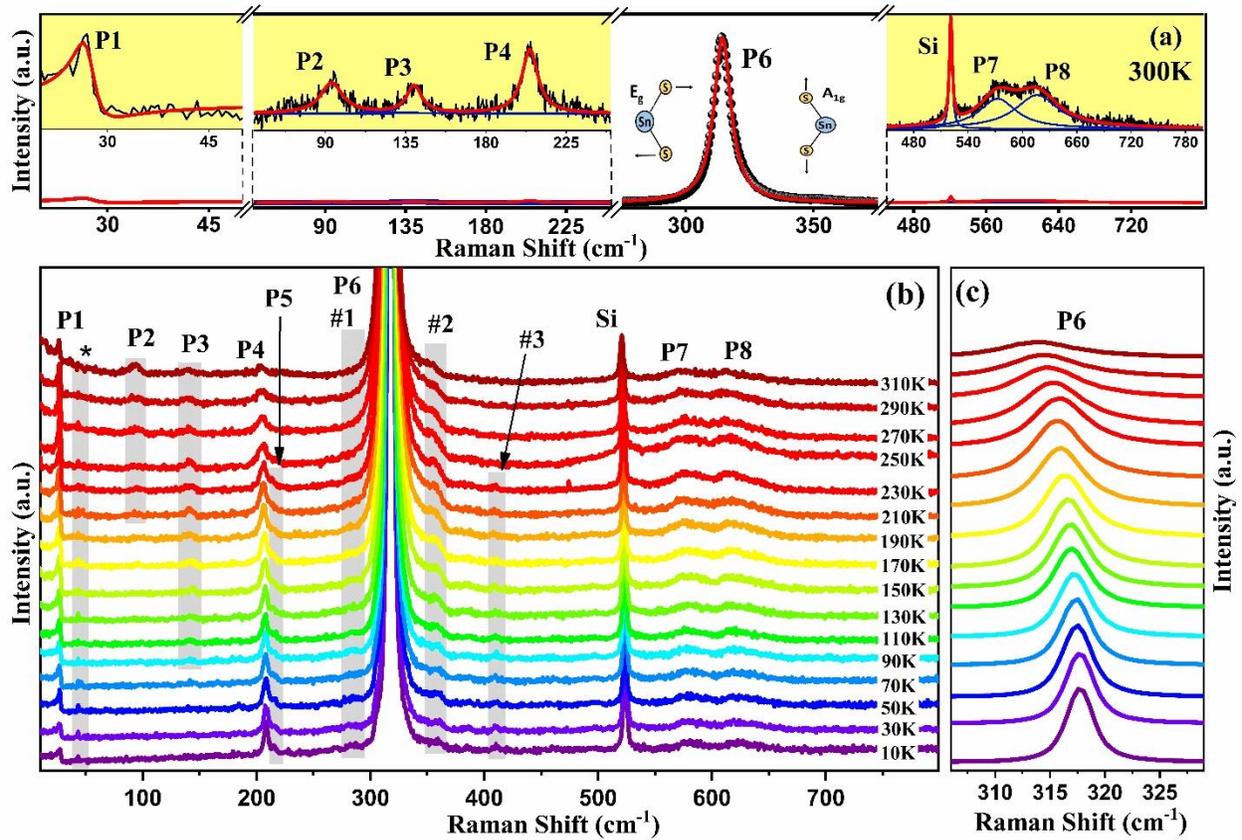

**Figure 1:** **(a)** Raman spectrum of the VA SnS$_2$ in a spectrum range of 10-800 cm$^{-1}$ recorded at 300 K. Insets in the yellow shaded area are the amplified spectra in the spectral range of 10 to 50 cm$^{-1}$, 50 to 250 cm$^{-1}$ and 450 to 800 cm$^{-1}$. Inset shows the ball-stick diagram representing the lattice vibration corresponding to the in plane ($E_g$) and out of plane ($A_{1g}$) Raman active modes. **(b)** Show the temperature evolution of Raman spectrum in the temperature range of ~ 10-310 K. **(c)** Shows the temperature evolution of the most prominent phonon mode P6 in the temperature range of ~ 10-310 K.

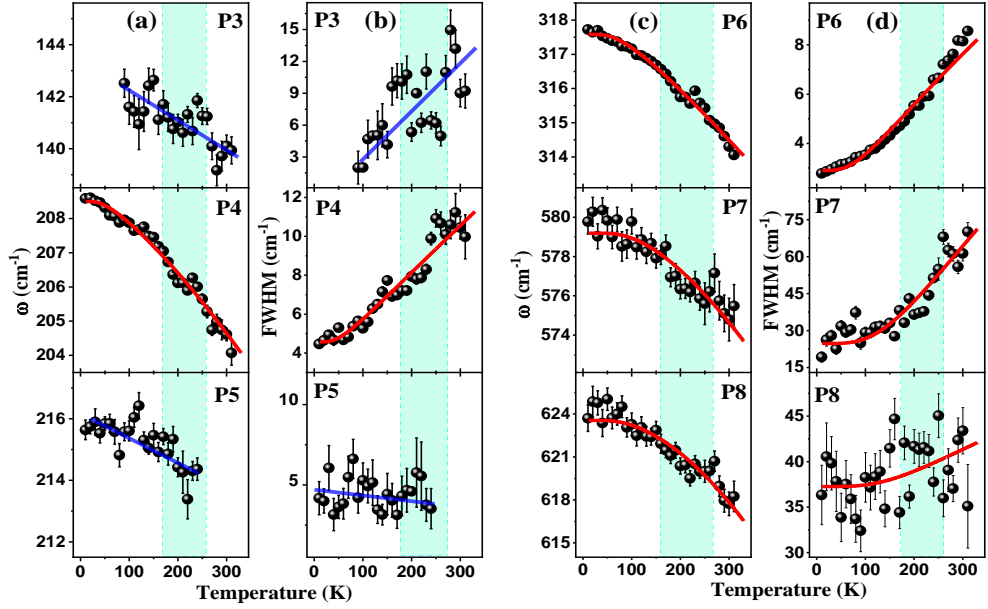

**Figure 2:** **(a)** and **(b)** Temperature dependence of frequency and FWHM of the phonon modes P3-P5 respectively; **(c)** and **(d)** Temperature dependence of the frequency and FWHM of the phonon modes P6-P8, respectively for VA $SnS_2$. The solid red lines are the fitted curves as described in the text and the semi-transparent blue lines are guide to the eye.

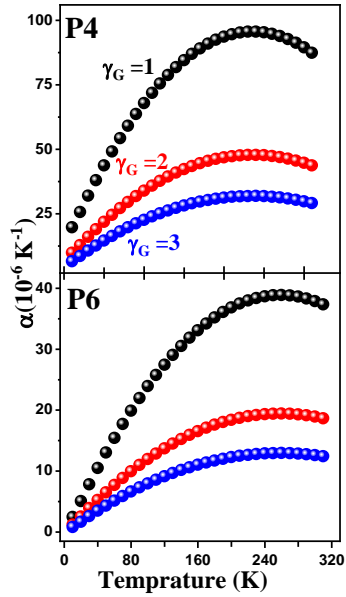

**Figure 3:** Show the variation of the linear thermal expansion coefficient as function of the temperature for the peaks P4 and P6 in the Temperature range of ~ 10-310 K.

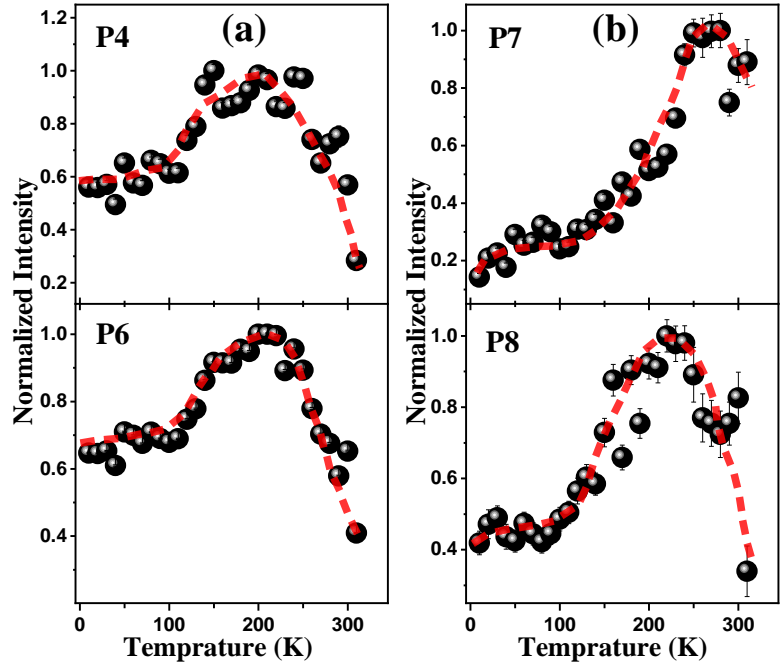

**Figure 4: (a)** and **(b)** shows the temperature dependence of the normalized intensity of the phonon modes P4, P6, P7, and P8, respectively. Dash red lines are drawn as guide to the eye.

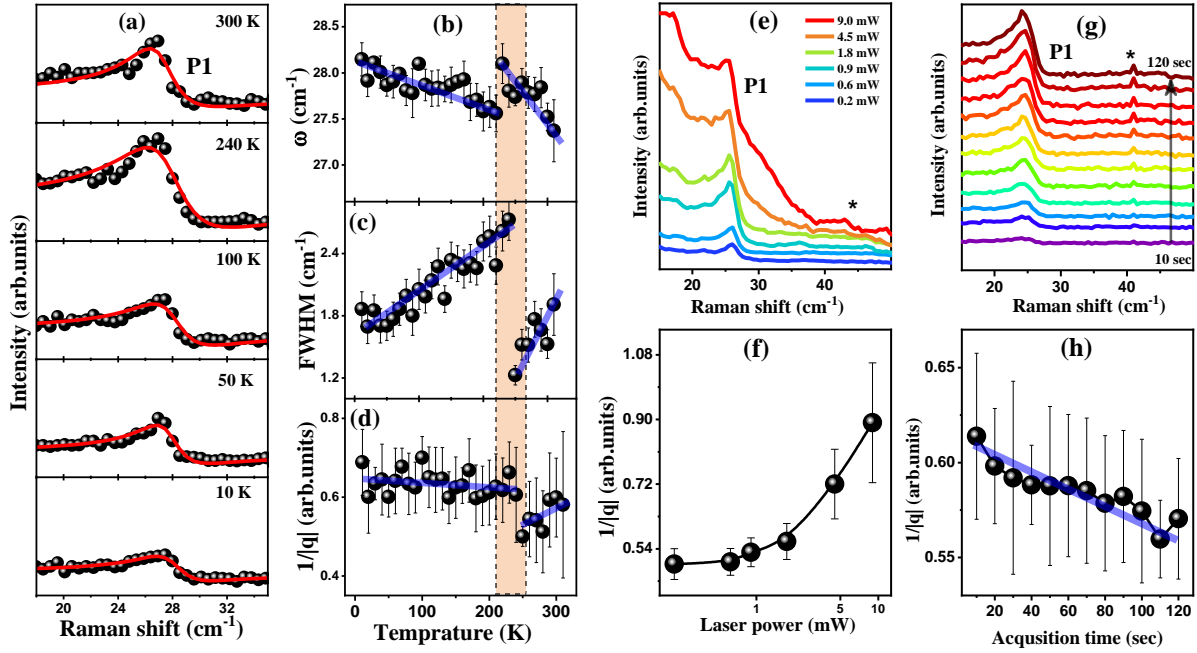

**Figure 5:** **(a)** Shows the temperature variation of the Fano peak (P1). **(b)** and **(c)** Show the temperature dependence of the frequency and FWHM of the Fano peak P1, respectively. **(d)** Temperature dependence of the coupling strength $1/|q|$. **(e)** Shows the laser excitation power dependence of the Raman spectrum in the frequency range of ~ 15 to 50 cm$^{-1}$ using laser excitation 532 nm. **(f)** Shows the Laser excitation power dependance of the coupling strength $1/|q|$ at room temperature. **(g)** Shows the evolution of Raman spectrum in the frequency range of ~ 15 to 50 cm$^{-1}$ as a function of spectra acquisition time at room temperature. **(h)** Shows the coupling strength $1/|q|$ as a function of the acquisition time. Semi-transparent blue lines are drawn as guide to the eye.

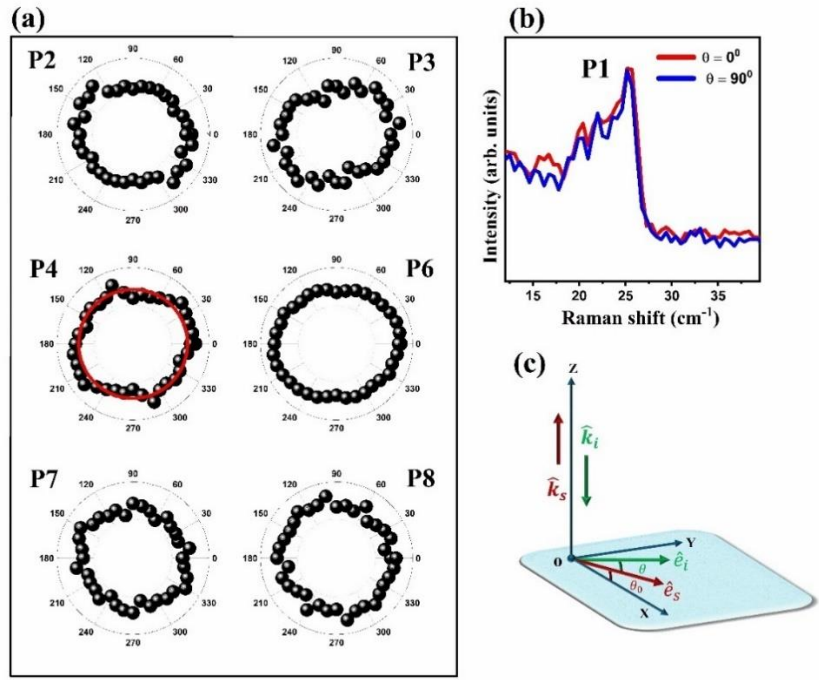

**Figure 6: (a)** Shows the intensity polar plot for VA SnS$_2$ using laser excitation 532 nm (2.33 eV) for peaks P2-P4, P6-P8 at 300K. **(b)** Shows the polarization spectra for the peak P1. **(c)** Shows the schematic for the polarization configuration.

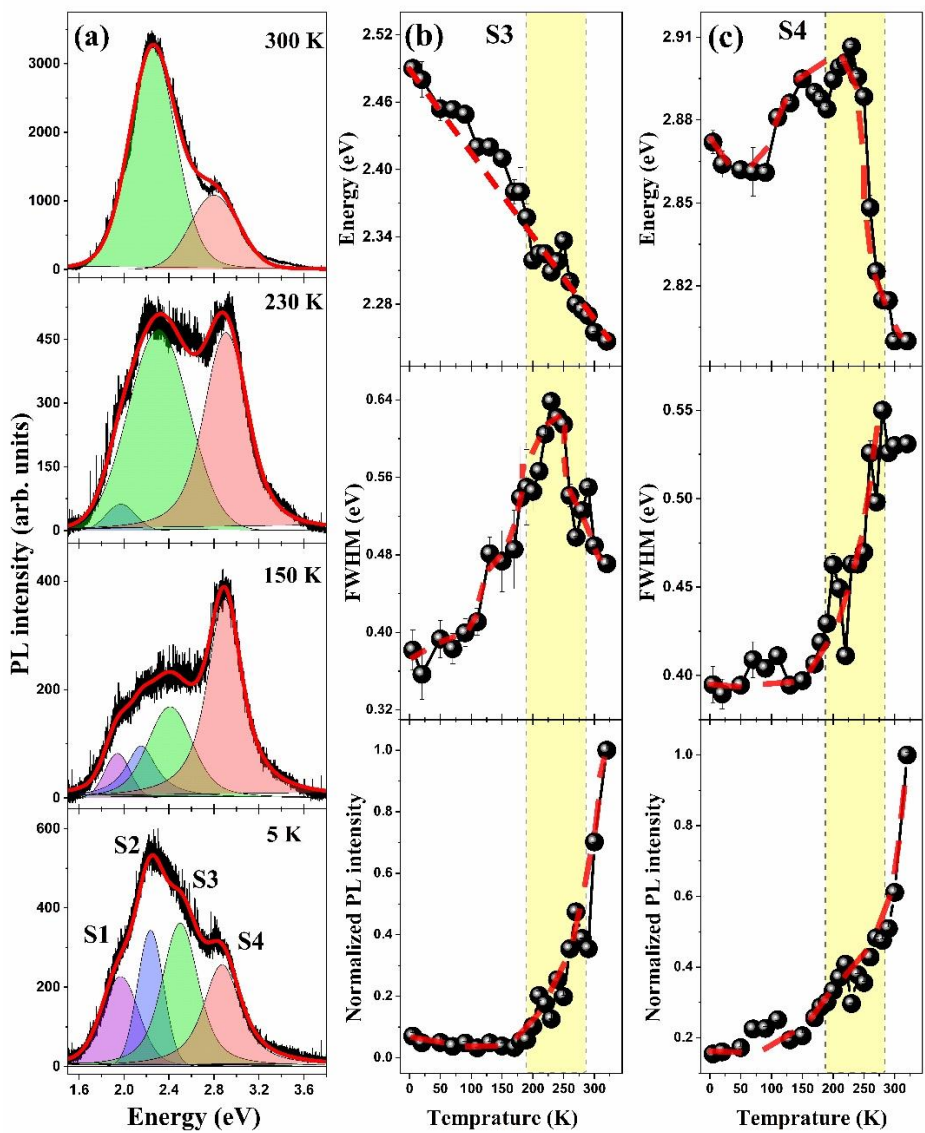

**Figure 7: (a)** Shows the evolution of the PL spectra. **(b)** and **(c)** Show the temperature dependance of the peak position, FWHM and intensity of the PL peaks S3 and S4, respectively. Dashed red lines are drawn as guide to the eye.

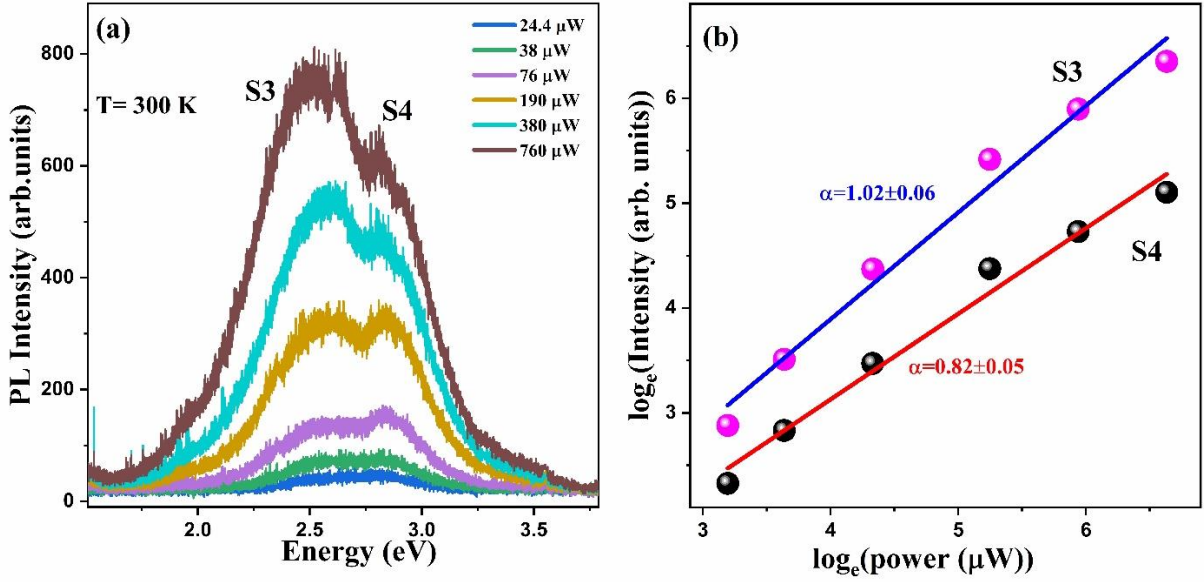

**Figure 8:** **(a)** Shows the evolution of PL spectra as function of power of laser excitation 325 nm (3.81 eV) at room temperature. **(b)** Shows the PL intensity of the peaks S3 and S4 as the function of the laser excitation (graph plotted on the $\log_e$-$\log_e$ scale). The solid lines are the fitted curve by power law: $I \propto P^\alpha$.

**Table I.** List of self-energy fitting parameters corresponding to the phonon modes in VA 2H-SnS$_2$, fitted using the equation as described in the text. The units are in cm$^{-1}$.

| Mode assignment | $\omega_0$ | A | B | $\Gamma_0$ | C | D |
|---|---|---|---|---|---|---|
| **P4** | 209.5±0.2 | -0.9±0.2 | 0.03±0.02 | 2.9±0.6 | 1.6±0.8 | 0.06±0.1 |
| **P6** | 319.3±0.3 | -1.7±0.4 | -0.02±0.1 | 1.5±0.1 | 1.2±0.1 | 0.2±0.1 |
| **P7** | 583.2±0.3 | -3.2±0.0 | -0.8±0.1 | 15.0±0.3 | 2.7±0.3 | 7.0±0.4 |
| **P8** | 626.1±0.8 | -0.6±0.2 | -2.0±0.2 | 35.5±0.2 | 1.0±0.3 | 0.7±0.3 |

Supplementary information:

**Tunable resonant Raman scattering with temperature in vertically aligned 2H-SnS$_2$**

Atul G. Chakkar[1, *], Deepu Kumar[1], Ashok Kumar[2], Mahesh Kumar[2], and Pradeep Kumar[1, #]

[1]*School of Physical Sciences, Indian Institute of Technology Mandi, Mandi-175005, India*
[2]*Department of Electrical Engineering, Indian Institute of Technology Jodhpur, 342037, India*

[*]E-mail: atulchakkar16@gmail.com
[#]E-mail: pkumar@iitmandi.ac.in

**Raman tensors for SnS$_2$:**

$$A_{1g} = \begin{pmatrix} a & 0 & 0 \\ 0 & a & 0 \\ 0 & 0 & b \end{pmatrix} \text{ and } E_g = \begin{pmatrix} c & 0 & 0 \\ 0 & -c & d \\ 0 & d & 0 \end{pmatrix}, \begin{pmatrix} 0 & -c & -d \\ -c & 0 & 0 \\ -d & 0 & 0 \end{pmatrix}$$

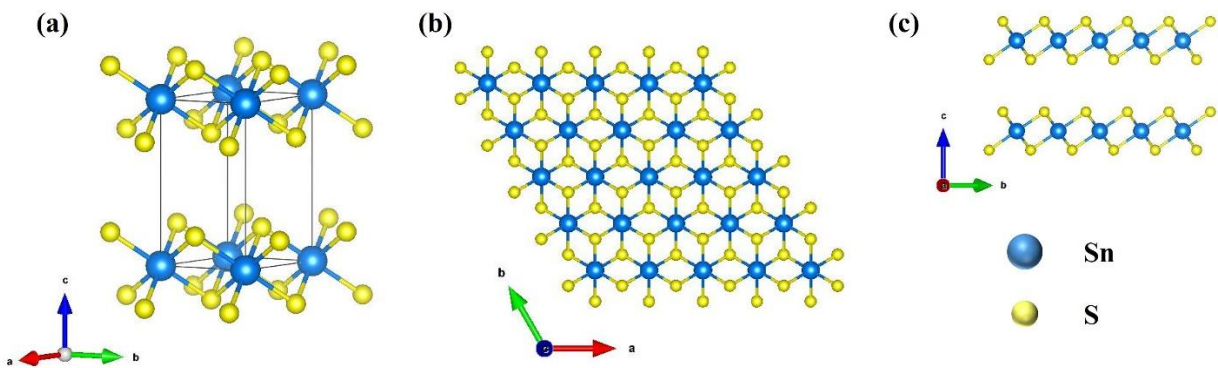

(a) (b) (c)

Sn
S

**Figure S1.** Shows the crystal structure of 2H $SnS_2$: **(a)** Shows the primitive unit cell. **(b)** and **(c)** Shows the top and side view of crystal structure.

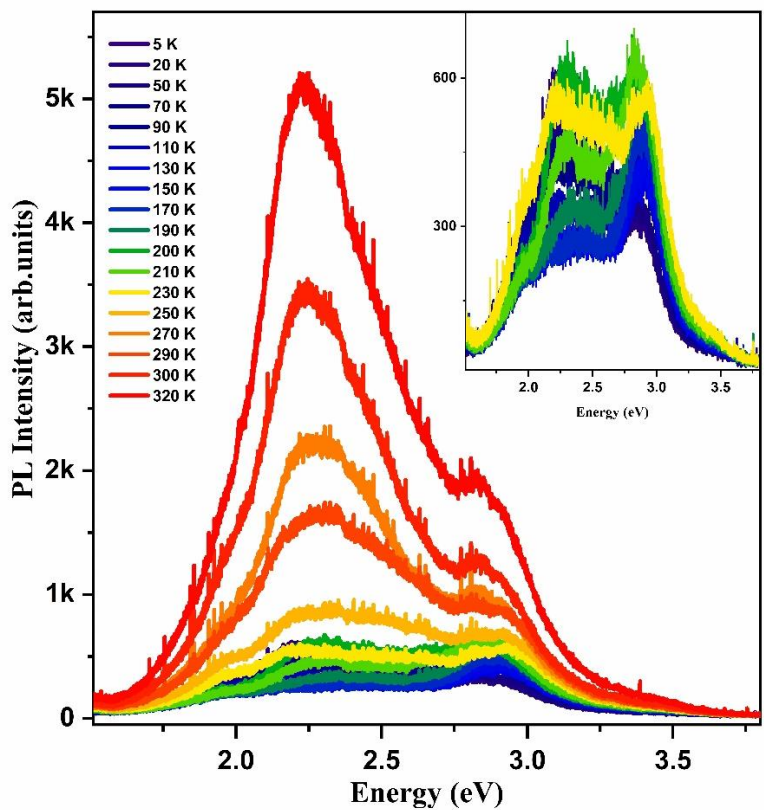

**Figure S2:** Shows the temperature evolution of the PL spectra for the VA 2H-$SnS_2$ in the temperature range of ~ 5-320 K. inset shows the evolution of the PL spectra for the low temperature range.